# Origin of the shadow Fermi surface in Bi-based cuprates

A. Koitzsch, S. V. Borisenko, A. A. Kordyuk, T. K. Kim, M. Knupfer, and J. Fink
*Leibniz-Institute for Solid State and Materials Research, IFW-Dresden, P.O. Box 270116, D-01171 Dresden, Germany*

M. S. Golden, W. Koops
*Van der Waals-Zeeman Institute, University van Amsterdam, Valckenierstraat 65, NL-1018 XE Amsterdam, The Netherlands*

H. Berger
*Institute of Physics of Complex Matter, EPFL, CH-1015 Lausanne, Switzerland*

B. Keimer and C. T. Lin
*Max-Planck Institute für Festkörperforschung, Heisenbergstrasse 1, D-70569 Stuttgart, Germany*

S. Ono and Y. Ando
*Central Research Institute of Electric Power Industry, Komae, Tokyo 201-8511, Japan*

R. Follath
*BESSY GmbH, Albert-Einstein-Strasse 15, 12489 Berlin, Germany*

**Abstract**: We used angle-resolved photoemission spectroscopy to study the shadow Fermi surface in one layer $Bi_2Sr_{1.6}La_{0.4}CuO_{6+\delta}$ and two layer $(Bi,Pb)_2Sr_2CaCu_2O_{8+\delta}$. We find the shadow band to have the same peakwidth and dispersion as the main band. In addition, the shadow band/main band intensity ratio is found to be binding energy independent. Consequently, it is concluded that the shadow bands in Bi-based HTSC do not originate from antiferromagnetic interactions but have a structural origin.

The Fermi surface and the electronic structure in general of the Bi-based high-$T_c$ cuprate family are among the most extensively studied objects in solid-state physics. By means of angle-resolved photoemission spectroscopy (ARPES), the intrinsic topology of the Fermi surface has been investigated in detail, as have the superconducting and pseudogaps and very recently even more subtle consequences of many body interactions on the electronic dispersion [1, 2]. In this context it is surprising that the so called shadow Fermi surface, a primal feature, which has been known of since the first angle-scanned Fermi surface maps of high temperature superconductors [3], has still not been understood thus far.

The shadow Fermi surface (SFS) appears, at a first glance, to be a shifted replica of the main Fermi surface. Its origin is controversial. Originally a magnetic origin was suggested by Aebi et al. motivated by earlier theoretical work [4]: electrons couple to short ranged antiferromagnetic fluctuations with the unit vector of the antiferromagnetic Brillouin zone ($\pi,\pi$). However, this interpretation has been cast into doubt [5] and a structural mechanism leading to a c(2x2) lattice superstructure component has been suggested [6]. Nevertheless the shadow Fermi surface sparked a profound theoretical effort – and continues to do so. In a series of studies the feature was found to be consistent [7] - or inconsistent [8] with antiferromagnetic fluctuations. On the experimental side the number of studies devoted to the shadow Fermi surface is limited. A previous ARPES investigation on $Bi_2Sr_2CaC_2uO_{8+\delta}$ (Bi-2212) found indications for the magnetic scenario by studying the energy dependence of the ratio of the amplitudes of main and shadow band [9]. Another ARPES study established that the SFS has a different intensity distribution to the corresponding sections of the main FS, which makes an explanation in terms of an extrinsic diffraction of the photoelectrons from the Cu-O bands on their way or through the surface of the crystal unlikely [10]. Based on the absence of any known intrinsic lattice superstructure, it was concluded in favor of a magnetic origin of the shadow Fermi surface. Two other studies could not support the magnetic scenario by investigating the polarization dependence [11] and the doping dependence [12]. An interesting effect was reported by Kordyuk et al.: it was found that the intensity ratio SFS versus main Fermi surface (MFS) shows a correlation with $T_c$ as a function of doping [13]. While this result seems incompatible with a magnetic scenario – for underdoping the intensity of the SFS is expected to increase because the antiferromagnetic order is approached – it is also not straightforwardly reconciled with a structural mechanism. Recently a VLEED (Very Low Energy Electron Diffraction) study revealed a hidden c(2x2) periodicity in pure Bi-2212, which is more consistent with a structural phenomenon [14]. Apart from Bi-2212, the SFS is also observed in the one-layer material $Bi_2Sr_{2-x}La_xCuO_{6+\delta}$ (Bi-2201) [15]. A feature referred to as a "shadow band" has been also reported for $Ca_{2-x}Na_xCuO_2Cl_2$ [15] and has been assigned a magnetic origin. However, the precise relationship of these shadow bands in the oxychloride to the the shadow Fermi surface in the Bi-based cuprates is uncertain at present.

We apply ARPES to study the spectral weight of main and shadow bands simultaneously as a function of binding energy for Bi-2201 and Bi-2212. The results are compared to the predictions of theory. We find clear disagreement in several aspects for the magnetic scenario. The results support instead a structural origin for the shadow bands in these systems.

The ARPES experiments were carried out using radiation from the U125/1-PGM beam line and an angle multiplexing photoemission spectrometer (SCIENTA SES 100) at the BESSY synchrotron radiation facility. The spectra were recorded using excitation energies $h\nu$=50-55 eV with a total energy resolution ranging from 10-30 meV. The momentum resolution was 0.01 Å$^{-1}$ parallel to (0,0)–($\pi,\pi$) and 0.02 Å$^{-1}$ perpendicular to this direction. Measurements have been performed on the one layer compound $Bi_2Sr_{1.6}La_{0.4}CuO_{6+\delta}$ and two layer $Bi_2Sr_2CaCu_2O_{8+\delta}$ high quality single crystals. The two-layer compound has been investigated in its pristine form and lead substituted. Lead is known to remove the (5x1) superstructure of the Bi-O layers, which gives rise to additional, well understood, diffraction replicas in pristine $Bi_2Sr_2CaCu_2O_{8+\delta}$ [17, 18]. At the same time the lead doping causes an additional c(2x2) superstructure [19]. In order not to confuse this Pb-c(2x2) superstructure with the possible origin of the shadow Fermi surface, data on pristine Bi-2212 are also presented.

Figure 1b shows a room temperature Fermi surface map of (Bi,Pb)-2212. The typical hole-like Fermi surface with barrels centered around the ($\pi,\pi$) and equivalent points is seen. It stems from the Cu-O derived bands cutting the Fermi energy and is qualitatively in agreement with LDA calculations based on a tetragonal undistorted unit cell yielding a closed Cu-O Fermi surface sheet around ($\pi,\pi$) and equivalent points [20]. The additional weaker barrel centered on the (0,0) and equivalent points is not predicted by these calculations. This is called the shadow Fermi surface. It can be viewed as a replica of the main barrels shifted by a ($\pi,\pi$) vector, the latter being the unit vector of the antiferromagnetic Brillouin zone. An electron at the Fermi surface involved in a scattering process with a momentum transfer ($\pi,\pi$) assumes a position on the shadow Fermi surface. This motivated the idea of the antiferromagnetic origin of this feature. The shadow Fermi surface is best separated from the main band along the (0,0)-($\pi,\pi$) direction. To study the properties of the shadow bands in comparison with the main band we performed "cuts" along (0,0)-($\pi,\pi$), i.e. we choose this direction as the momentum axis and recorded the dependence on binding energy. These "energy distribution maps" (EDM's) are shown in Figures 1(a, c, d) for different samples. Figure 1a depicts the EDM for an underdoped superstructure-free (Bi,Pb)-2212 sample with $T_c$= 76 K.. For a quantitative treatment we fitted horizontal cuts of this two-dimensional data set (momentum distribution curves, or MDC's) with two Lorentzians. The loci of the maxima of these Lorentzian peaks are marked in the figure and denote the experimental dispersion relation of the feature under consideration. Panel c shows an equivalent dataset for a lead-free (i.e. pristine) Bi-2212 sample ($T_c$= 85 K, underdoped). In this case additional features appear in between the shadow and main band. These are the diffraction replicas of the main and shadow bands, typical of pristine Bi-2212 [17, 10]. In panel (d) an EDM from one-layer, lead-free Bi-2201 ($T_c$=32 K) is presented. Shown are the shadow band (left) and the diffraction replica of the main band (right). In the following we adopt the strategy to single out certain properties of the electronic states which an antiferromagnetic scattering process would impose and to compare these expectations with the experimental data.

In Fig. 2 b, d, f the momentum widths of the main bands (diffraction replica of the main band for Bi-2201) and shadow bands are compared as a function of binding energy. The coupling of the fermions to overdamped antiferromagnetic fluctuations would imply a broadening of the shadow bands in both momentum and energy. Since the correlation length for antiferromagnetic fluctuations is known from neutron scattering and NMR experiments to be only a few lattice constants [21, 22], the momentum broadening should be quite severe: of the order of 0.1 Å$^{-1}$. As can be seen from Fig. 2, the observed widths are similar and almost identical within the statistical errors of the fit for all binding energies and all samples. The remaining minor discrepancies could be the result of a slight difference in alignment of the k-space cut. In any case, there is no indication of momentum broadening of the shadow band.
A magnetic scattering channel would also entail an energy renormalization of the shadow band states compared to the main band, since the scattering process involves an energy transfer. Within this view, the shadow band can be viewed as satellite line of the main band. Below $T_c$ the spin fluctuation spectrum of optimally doped Bi-2212 is sharply dominated by the magnetic resonance at $(\pi,\pi)$ vector with an energy $\omega_0$=43 meV and 10-15 meV width [23]. $\omega_0$ is expected to be slightly lower for the slightly underdoped Bi-2212 samples used here. Nevertheless the satellite intensity should set in approximately at $\omega_0$ below the Fermi energy. At the Fermi energy no spectral weight is expected if the energy resolution is better then $\omega_0$ as is the case here. Furthermore, for all energies $\omega>\omega_0$ the main contribution of the shadow band comes from the main band states shifted by $\omega_0$. Therefore, the dispersion of main and shadow band should clearly differ. Fig. 2 a, c, e compare the dispersions of the shadow and main bands (diffraction replica of the main band for Bi-2201). We find good agreement and no sign of systematic deviations. There is virtually no renormalization between shadow and main band.

It has been pointed out previously that the intensity ratio between shadow and main band should depend on binding energy if the shadow band is due to antiferromagnetic fluctuations [9]: the ratio must tend to zero at the Fermi energy as is clear from the above discussion. If the shadow band is a structural replica of the main band, i.e. is due to a static potential, a constant ratio is expected. Figure 2 b, d, f (insets) show the results for the intensity ratio for the three samples considered. We define the intensity here as the area under the peak as extracted from the Lorentzian fits described above. No significant dependence of the ratio on binding energy is apparent in Fig. 2 b, d, f (insets). Evaluating this ratio requires rather good statistics. For the one layer compound in panel (f) the ratio between the shadow band and the first diffraction replica of the main band is depicted. This is justified if the ratio between diffraction replica and main band is also constant as will be shown below. In contrast, in a previous paper [9] a strong energy dependence of the amplitude ratio in rough agreement with theoretical predictions for the magnetic scenario was found. However, this result is based on an EDC analysis (energy distribution curve, intensity at essentially constant momentum as a function of energy) where the background subtraction is decisive and thus it is very difficult to evaluate intensities extracted in this way quantitatively. Moreover the density of k-points used in the measurements presented here is ca. 10 times greater than that used in Ref. [9].
We additionally investigated the temperature dependence of the shadow band/main band ratio (SB/MB) at $E_F$ for one of the samples (Fig. 2a inset). Since the correlation length for antiferromagnetic order is known to depend on temperature, a temperature dependence of the ratio would be expected if the shadow bands were due to coupling to antiferromagnetic spin fluctuations. Although not numerous, the datapoints in Fig. 2a do not support such a conjecture: a constant ratio is observed within error bars.

To summarize the situation thus far: we did not observe any signatures of a significant influence of antiferromagnetic fluctuations on the shadow band, either in pristine or Pb-doped Bi-2212 nor in pristine Bi-2201. The question then naturally arises as to what else the origin of the shadow band may be?

In Fig. 2 c,e (insets) we present a comparison of the dispersion and the intensity ratio for the main band and the first diffraction replica for the pristine Bi-2212 sample. As expected we find quantitative agreement for the dispersion and a constant intensity ratio as a function of binding energy. The full width half maximum at $E_F$ is $(0.029\pm0.001)$ Å$^{-1}$ for the main band and $(0.033\pm0.002)$ Å$^{-1}$ for the diffraction replica, which is again in agreement. These are analogous results as we obtained for the SB-MB comparison for the same sample. By analogy we would thus be led to conclude that the shadow band itself is another type of replica. Hence it must have a structural origin.

It has been shown that LDA calculations, taking into account the precise orthorhombic unit cell along with lattice distortions, rather than resorting to the tetragonal unit cell as usually done, result in a additional Fermi surface barrel centered at (0,0), which is essentially a back-folding of the bandstructure due to the doubling of the unit cell [6]. Such a unit cell doubling would for instance occur if every second copper atom in the Cu-O plane is structurally inequivalent, due to a hidden distortion of the lattice, e.g. a buckling. This scenario, which would lead to changes in both the occupied (initial) and unoccupied (final) states in the photoemission experiment would be fully consistent with the above findings.

Another structural explanation would be to assume that the outgoing photoelectrons (from an undistorted, tetragonal $CuO_2$ plane) are subsequently diffracted by a hidden c(2x2) structure of the block layers. However, the shadow bands are observed to be essentially identical in lead-free and lead-doped samples. It seems unlikely that the lead doping efficiently suppresses the incommensurate (5x1) superstructure in the Bi-O layers without interfering with a possible c(2x2) feature outside the $CuO_2$ planes. Therefore, the conclusion that the shadow bands most likely originate from the Cu-O layers and are not the result of an extrinsic diffraction of the sort which gives rise to the incommensurate super-structure induced 'diffraction replicas' seems very reasonable.

The structural scenario is supported by the recent observation of a c(2x2) superstructure in pure Bi-2212 using VLEED [14]. But not all questions are settled: the previously reported dependence of the SFS/MFS intensity ratio on doping in the vicinity of the ($\pi$,0) point remains puzzling [13]. We speculate that small structural changes may occur when the as-grown crystals, which are about optimally doped, are annealed in vacuum/argon or oxygen to achieve under- or overdoping respectively.

To summarize, we have studied the shadow and main bands for k along the (0,0)-($\pi$,$\pi$) direction in lead doped and lead-free Bi-2212 and lead-free Bi-2201. We found that the width of the MDC's and band dispersions were essentially identical between the main and shadow bands. In addition, the SB/MB intensity ratio was found to be independent of binding energy. Furthermore, no significant temperature dependence of the SB/MB ratio was observed. These findings are inconsistent with a scenario where the shadow band is attributed to scattering due to short-ranged, overdamped antiferromagnetic fluctuations. Assuming that these data for the nodal direction are representative for the rest of the Brillouin zone, one is thus led to conclude that the shadow Fermi surface in Bi-based cuprates has a structural origin.

We thank R. Hübel for technical support and K. Nenkov for the $T_c$-measurements. MSG and WK are grateful to FOM for support. The work in Lausanne was supported by the Swiss National Science and by the MaNEP.

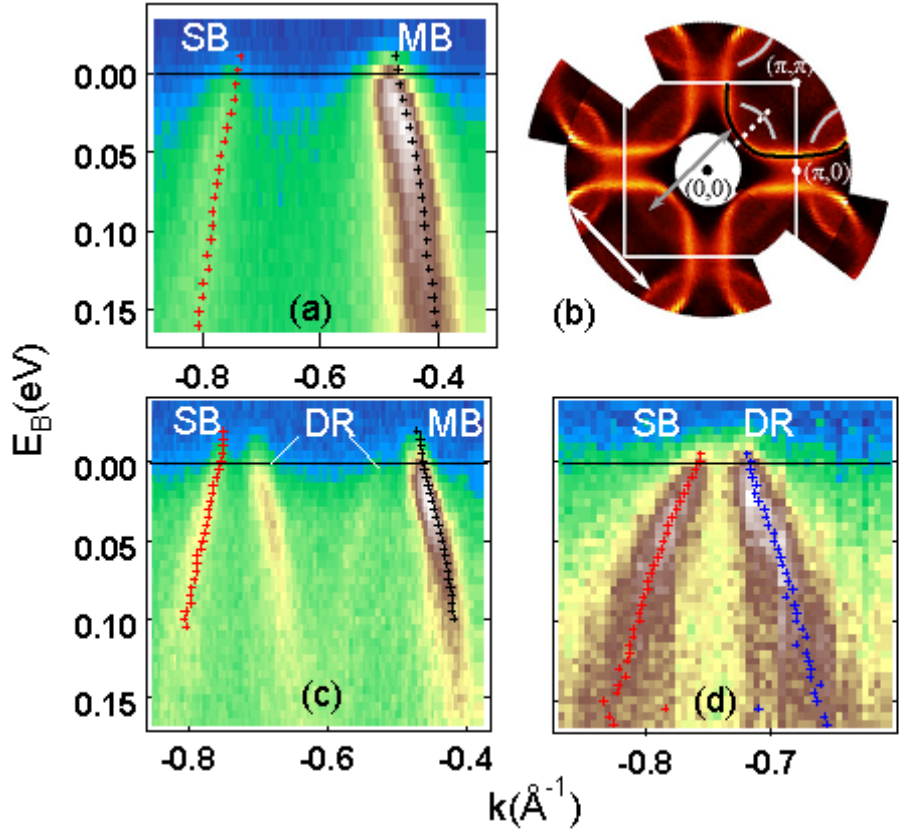

**Figure 1** (Color): (a), (c), (d) Energy distribution maps (EDM) along the (0,0)-(π,π) direction for (a) underdoped-(Bi,Pb)-2212, taken with excitation energy 55 eV at T=30 K; (c) slightly underdoped lead-free Bi-2212, excitation energy 50 eV at T=30 K (d) optimally doped Bi-2201, excitation energy 50 eV at T=65 K. Main bands (MB), shadow bands (SB) and diffraction replicas (DR) are indicated. The markers denote the dispersion as yielded by a fit (see text). (b) Fermi surface map of an overdoped (Bi,Pb)-2212 ($T_c$=69 K) crystal at room temperature taken with excitation energy 21.2 eV [13]. The white square represents the first Brillouin zone. The main Fermi surface is emphasized by a black line in the upper right part, the shadow Fermi surface with white. The dashed line marks the regions where the EDM's have been taken. The arrows represent (π,π) and equivalent vectors.

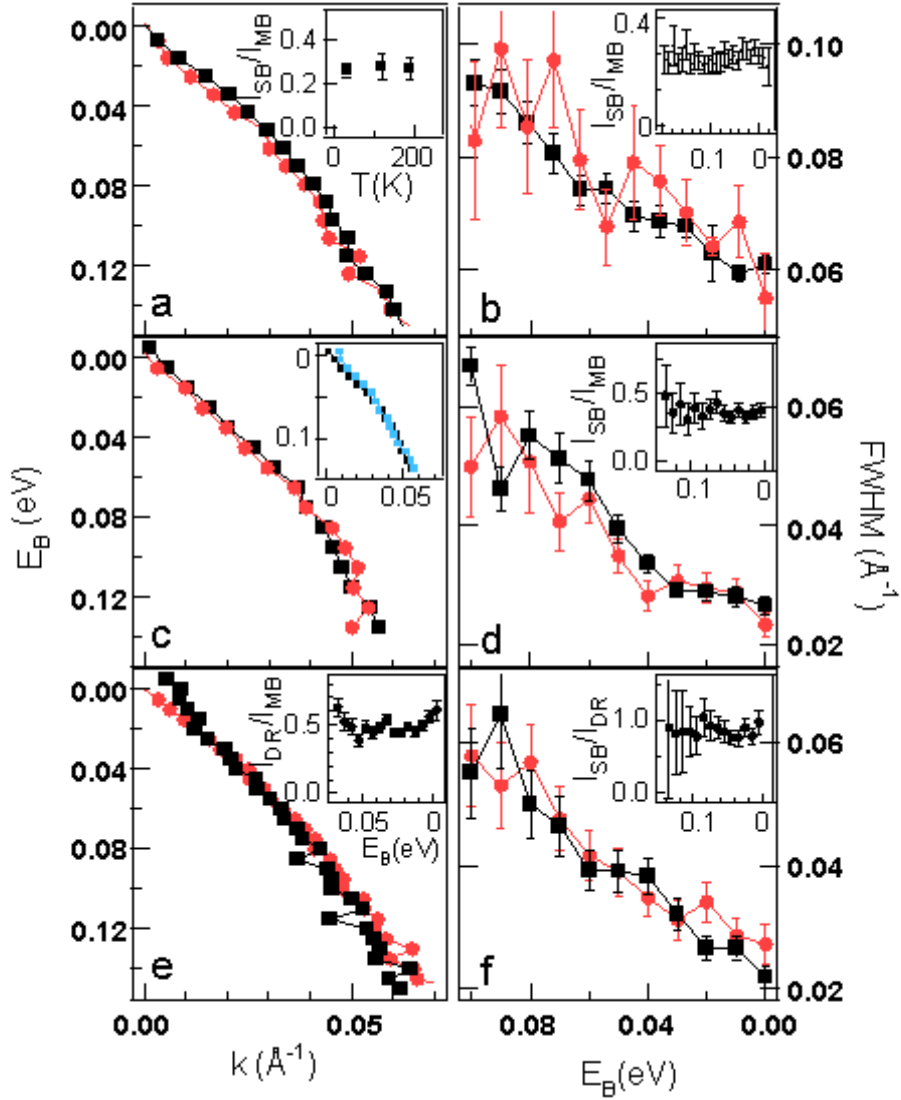

**Figure 2** (color): Comparison of the dispersions of SB (light circles) and MB or DR (dark squares) shown in Fig. 1 and corresponding peakwidths for (a, b) underdoped (Bi,Pb)-2212; (c, d) slightly underdoped Bi-2212 and (e, f) optimally doped Bi-2201. The dispersion of the SB have been mirrored and the curves have been shifted for better comparison. The MDC peakwidths in the (Bi,Pb)-2212 data are a bit broader than the other data due to different resolution settings in this case. (Inset c) shows a comparison of the dispersions of MB (dark) and DR (light) for the pure Bi-2212 sample. (Inset a) shows the ratio of the peak areas of SB and MB as a function of temperature and (Inset b) as a function of energy for underdoped (Bi,Pb)-2212. (Inset d): energy dependent intensity ratio for slightly underdoped Bi-2212. (Inset e): energy dependent intensity ratio of the peak areas of DR and MB for the pure Bi-2212 sample. (Inset f): energy dependent intensity ratio of the peak areas of SB and DR for the Bi-2201 sample.